\documentclass[authoryear,review,3p,10pt,onecolumn,leqno]{elsarticle}
\usepackage[english]{babel}
\usepackage[utf8]{inputenc}

\usepackage{caption}
\usepackage{subcaption}
\usepackage{todonotes}

\usepackage{epsfig, epstopdf}
\usepackage{color}
\usepackage{hyperref}
\usepackage{lineno}
\usepackage{amssymb,amsmath}
\usepackage{natbib}
\usepackage{setspace}
\usepackage{lscape}
\usepackage{afterpage,latexsym,epsfig,float}
\bibpunct{(}{)}{;}{a}{}{,}

\def\by{\mathbf{y}}

\def\bu{\mathbf{u}}
\def\bK{\mathbf{K}}
\def\bepsilon{{\boldsymbol \epsilon}}

\journal{Preprint version}

\begin{document}

\title{Comment on ``A variational Bayesian approach for\\ inverse problems with skew-t error distributions"\\ (Guha et al.,
J. Comput. Phys. 301 (2015) 377--393)}

\author[IFOP,UTFS]{Javier E. Contreras-Reyes\corref{cor1}}
\ead{jecontrr@uc.cl, javier.contreras@ifop.cl}
\author[UTFS]{Freddy Omar L\'opez Quintero}

\cortext[cor1]{Corresponding author. Phone +56 032 2151682, Fax +56 032 2151645}
\address[IFOP]{Divisi\'on de Investigaci\'on Pesquera, Instituto de Fomento Pesquero, Blanco 839, Valpara\'iso, Chile}
\address[UTFS]{Departamento de Matem\'aticas, Universidad T\'ecnica Federico Santa Mar\'ia, Valpara\'iso, Chile}

\begin{abstract}
A brief comment on {\it A variational Bayesian approach for inverse problems with skew-t error distributions} (Guha et al., Journal of Computational Physics 301 (2015) 377-393) is given in this letter.\\
\end{abstract}

\maketitle

\vskip5mm

First of all, we want to thank \citet{Guha_et_al_2015} for their paper. They proposed a variational Bayesian approach to inverse problems with skew-$t$ distributed errors. Their paper considers two finite-dimensional linear inverse problems: the Cauchy--Laplace equation and the Multi-phase flow. For each one,
the following system with additive noise is considered:
\begin{eqnarray}
\by=\bK(\bu) + \bepsilon,\label{A0}
\end{eqnarray}
\noindent where $\bK(\bu)$ corresponds to the model output from the forward model $\bK$, $\bu$ is the solution, and $\bepsilon$ is the additive error.
 According to Bayes' theorem, the posterior distribution $p(\bu|\by)$ of the unknown $\bu$ is $p(\bu|\by)\propto p(\by|\bu)p(\bu)$. Assuming that each noise component $\epsilon_i$ is independent and skew-$t$ identically distributed, its density function is
\begin{eqnarray}
p(\epsilon_i|\sigma^2,\alpha,\nu)\propto \frac{1}{\sigma}t(\frac{\epsilon_i}{\sigma};\nu)
T\left(\frac{\alpha\epsilon_i}{\sigma}\sqrt{\frac{\nu+1}{\nu+(\epsilon_i/\sigma)^2}};\nu+1\right),\label{A1}
\end{eqnarray}
\noindent where $t(\cdot;\nu)$ is the symmetric Student-$t$ density with $\nu$ degrees of freedom and $T(\cdot; \nu+1)$ its respective cumulative distribution function, $\sigma$ a scale parameter, and $\alpha$ a skewness parameter. Then, for each $\epsilon_i=y_i-\bK(\bu)_i$, the following hierarchical model \citep{Cancho_et_al_2011} is proposed:
\begin{eqnarray}
\epsilon_i & = & \Delta z_i +w_i^{-{\frac{1}{2}}} \tau^\frac{1}{2} N_i \label{A2}
\end{eqnarray}
\noindent where $M_i$ and $N_i$ are independent and standardized normal identically distributed, $\tau=\sigma^2(1-\delta^2)$, $\delta=\alpha/\sqrt{1+\alpha^2}$, $w_i\sim\mbox{Gamma}(\nu/2,\nu/2)$, $\Delta=\sigma \delta$ and $z_i=w_i^{-{\frac{1}{2}}}|M_i|$.

In this letter, we focus on the error distribution of the additive noise. We believe that posterior distribution \citep[Eq. (6) of ][]{Guha_et_al_2015}, corresponding to the full Bayesian solution to the inverse problem (\ref{A0}), does not encapsulate all the information about the problem. The authors mentioned they take advantage of the stochastic representation (\ref{A2}) to establish their likelihood function. However, we noticed that, even if valid, their likelihood does not include the explicit representations of $\mathbf{z}$, $\mathbf{w}$ or $\nu$, as in Eq. (8)--(10) of \citet{Cancho_et_al_2011}. It can be noted that all these parameters are part of representation (\ref{A2}).

As in a previous work of one of the authors, \citet{jin_2012}, an interesting approximation was developed that avoided the inclusion of $\nu$ and $\sigma$ parameter of the $t$-distribution, assuming these do not accept the easy conjugate form and could be estimated via maximum likelihood when appropriate. On this particular, \citet{wand_et_al_2011} have developed explicit results in variational Bayesian analysis. Nevertheless, our main concern is that the variability that could be captured by the $\nu$ parameter, as we understand, has been reduced to an assumed value. \citet[][page 454]{gelman_et_al_2004}, resolves this issue adding a Metropolis step just by sampling the $\nu$ parameter.


\citet{Geweke_1993} and \citet{Cancho_et_al_2011}, for example, consider as prior distribution a truncated exponential density located in the interval $(2,\infty)$ with mean $0.5$, denoted as $TE_{(2,\infty)}(0.5)$. This means that $\nu$ should be strictly larger than 2 to ensure the existence of variance in the skew-$t$ model. Besides, $\nu$ is directly estimated from the posterior density, whereas in the classical approach used by \citet{Lange_Sinsheimer_1993}, it is obtained manually using profiles of the log-likelihood function. This prior distribution is also used by \citet{Contreras_et_al_2015}. They concluded that associated error parameter estimates for non-linear von Bertalanffy growth function are smaller than parameters estimated by the classical approach. Basically, they determined: (i) the log-skew-$t$ (a slow variant of skew-$t$ model with positive values of $y_i$) is the best model among all the competing models (log-skew-normal, log-Student-$t$, and log-normal) using a WAIC as an information criteria for selecting models; (ii) the shape parameter $\alpha$ and degrees of freedom parameter $\nu$ provide a better solution in terms of confidence error of parameter estimates; and (iii) the advantage of the Bayesian approach is the feasibility to estimate the $\nu$ parameter directly, instead of using profiles of the log-likelihood function or just assuming $\nu$ is known.

The authors state: ``Due to the presence of several hyper-parameters and the intractable normalizing constant, the posterior distribution of Eq. (6) is not
explicitly available in closed form. One way to explore the high-dimensional posterior state space is to use MCMC based methods to simulate samples from
the posterior distribution. However, it is well known that the convergence of the chain is often not easy to diagnose". We agree only partially. True, convergence issues are challenging to diagnose for all approximate methods. For example, in \citet{Contreras_et_al_2015}, chains
were visualized throughout trace and autocorrelation plots. Geweke, Heidelberger--Welch,  and effective sample size (ESS) tests \citep{Cowles_Carlin_1996}
were used for this proposal. To simulate the chains for log-skew-t model, $2\times10^4$ iterations were taken.

Because the presented variational Bayes is highly competitive in computational time terms, we are sure that including this step will not represent a notable slowing of the algorithm. In addition, the impact of not having considered the complete \citet{Cancho_et_al_2011} representation is a not easy to quantify and deserves a careful sensitivity analysis.

To conclude, we encourage Guha et al. to take into account the considerations of our comment, mainly incorporating the prior specification claimed here (or elsewhere) in posterior distribution and completing the stochastic representation. Moreover, our approach can explain the reason why it is not efficient to obtain all the information about the inverse problem, such as possible converge problems.

\end{document}